\documentclass[a4paper,11pt]{article}

\usepackage{jcappub}

\usepackage[T1]{fontenc}
\usepackage{multirow}
\title{An improved upper limit to the CMB circular polarization at large
angular scales}

\author[a]{R. Mainini,}
\author[a,1]{D. Minelli,\note{now at Institute of Plasma Physics of the
 Italian National Research Council, IFP-CNR, Milano}}
\author[a]{M. Gervasi,}
\author[a]{G. Boella,}
\author[a]{G. Sironi,}
\author[a]{A. Ba\'u,}
\author[a]{S. Banfi,}
\author[a]{A. Passerini,}
\author[a]{A. De Lucia,}
\author[b]{F. Cavaliere}

\affiliation[a]{Physics Department, University of Milano Bicocca, \\Milano,
I20126}
\affiliation[b]{Physics Department, University of Milano, \\Milano, I20133}
\emailAdd{roberto.mainini@mib.infn.it}

\begin{abstract}
{Circular polarization of the Cosmic Microwave Background (CMB) offers the
possibility of detecting rotations of the universe and magnetic
fields in the primeval universe or in distant clusters of galaxies. We used the Milano
Polarimeter (MIPOL) installed at the Testa Grigia Observatory, on the italian Alps,
to improve the existing upper limits to the CMB
circular polarization at large angular scales. We obtain $95\%$ confidence level 
upper limits to the degree of the CMB circular polarization ranging
between $5.0 \cdot 10^{-4}$ and $0.7 \cdot 10^{-4}$
at angular scales between $8^{\circ}$ and $24^{\circ}$, improving
by one order of magnitude preexisting upper limits at large angular scales.
Our results are still far from the
$nK$ region where today expectations place the amplitude of the $V$ Stokes
parameter used to characterize circular polarization of the CMB but improve
the preexisting limit at similar angular scales.
Our observations offered also the opportunity of characterizing
the atmospheric emission at 33 GHz at the Testa Grigia Observatory.}

\end{abstract}

\begin{document}
\maketitle
\flushbottom

%\keywords{diffuse radiation, galactic emission, cosmic microwave
%background, unresolved extragalactic sources, sky absolute
%temperature}

\section{Introduction} \label{intro}
Polarization of the Cosmic Microwave Background (CMB) is a second order effect of matter radiation
interactions during the Universe evolution. Linear polarization, produced by
Thomson scattering of the CMB on matter anisotropies at the last scattering surface, has been detected
at the $\mu K$ level at various angular scales (e.g. \cite{DASI, WMAP, Boom}). Linear
polarization can be produced also if CMB interacts with primordial gravitational waves. Detecting
this component, known as B-modes of linear polarization and expected at the $nK$ level or
below, is extremely challenging. It is the aim of various experiments in preparation (e.g. \cite{Qubic, lspe}) .
\par The only attempts so far performed of detecting circular polarization
of the CMB were made in the'80s when
the search for fine structures of the CMB started (see Table \ref{Ta1}).
Circular polarization
was in fact searched as a possible signature of non uniform expansion and
rotation of the Universe which characterize some Bianchi Models
\cite{negroponte, basko, stark, tolman}.
The search however was abandoned when non uniform expansion
and rotation of the Universe were not supported by other observations
(e.g. \cite{sloan}). But interest to constraining anisotropic expansion
of the Universe is still present \cite{cai} and vorticities associated
to Bianchi VII Cosmology have been strongly constrained but not completely excluded
by the most recent CMB observations \cite{bianchi}.
\par Other processes which may induce circular
polarization of the CMB were then considered (e.g. \cite{cooray,bavarsad}
and references therein). For instance
circular polarization is expected beside linear
polarization when the CMB Thomson scattering occurs in presence of
background magnetic fields, (see \cite{giovannini2010} and references therein), 
or in  weakly
magnetized plasmas \cite{giovannini1997} and appears everytime the
photon scattering is completely forward \cite{sawyer}.
\par The expected amplitude of these circularly polarized signals is
very faint, probably $\leq nK$, not very different from the expected
amplitude of the B-modes linear polarization.
But while experiments for detecting B-modes are currently
underway no other experiment aimed at detecting circular
polarization of the CMB at the same
level or even at higher level has been proposed. The upper limits
obtained by \cite{lubin} and  \cite{partridge} (see Table \ref{Ta1})
in the '80 are still the only results one can find in literature.
\par In view of the information they can provide
it seems now time for new attempts of detecting CMB circular
polarization with sensitivities  at $nK$ level, but
a few years will be necessary before they will be ready. So, while
waiting for them we decided to exploit the almost
unique capability of MIPOL, among the existing CMB instrumentation,
and analyzed data we collected in 2009-2010.

% ************* Osservazioni in letteratura
\begin{table}
\begin{center}
\begin{tabular}{|ccccc|}
\hline
Reference&Wavelength&Angular scale&Polarization degree& Sky region\\
&$\lambda$ (cm)&$\Delta \theta$&$\Pi_V$&\\
\hline
\cite{lubin}&0.91&$15^{\circ}$&$\leq 4 \cdot 10^{-3}$&$\delta=+37$\\
\cite{partridge}&6.0&$18'' - 160''$&$\leq (2.2 - 0.6) \cdot
10^{-4}$&$\delta=+80$\\
\hline
\end{tabular}
\caption{Summary of CMB circular polarization upper limits at 95\% CL: $\Pi_V = V/T^{CMB}$.}
\end{center}
\label{Ta1}
\end{table}
%++++++++++++++++++++++++++++++++++++++

\section{MIPOL : Milano Polarimeter } \label{mipol}
\par Let's assume a radiation flux of brightness temperature $T$,
mixture of polarized (temperature $T_p$)
and unpolarized (temperature $T_{up}$) radiation. We can write
$T_p = \sqrt{U^2+Q^2+V^2~}$ where $U$, $Q$ and $V$ are
the so called Stokes Parameters: $U$ and $Q$ describe
the linearly polarized component of $T_p$, $V$ the
circularly polarized component, (see
\cite{stokes} and references therein).
\par MIPOL (Milano Polarimeter) is a 33 GHz ($\lambda =$ 9.1 mm) two channel,
$(0-\pi)$ phase modulated, etherodyne
correlation receiver \cite{siro98}, \cite{spiga}, \cite{spie}. From
$T$ MIPOL extracts a pair of Stokes Parameters of
the radiation which hits the antenna:
$V$ and $U$ ({\it Circular} or $C-$mode) or $Q$ and $U$
({\it Linear} or $L-$mode).
The antenna, a corrugated horn with an orthomode transducer, equipped with an
apodized ground shield rigidly attached to the horn, has a $14^{\circ}$ FWHM beam
which can be reduced to $7^{\circ}$ adding a proper extension to
the horn mouth. The orthomode transducer splits the total
(polarized and unpolarized) incoming radiation of temperature $T$ in two
linearly polarized components with crossed electric vectors
${\bf E}_1$  and ${\bf E}_2$ (temperatures $T_1 \propto E_1^2$ and
$T_2\propto E_2^2$) which then propagate through different channels
$Ch_1$ and $Ch_2$.
\par\noindent After proper amplification and coherent
frequency conversion the two signals go to:
\par i) total power detectors whose outputs
\begin{eqnarray}
TP_1 = S_{TP_1}T_1 \nonumber \\
TP_2 = S_{TP_2}T_2 \label{eq:TPi}
\end{eqnarray}
\par\noindent monitor the antenna temperatures $T_1$ and $T_2$
produced by polarized and unpolarized components of the sky signal plus system noise;
\par ii) a phase discriminator whose outputs
\begin{eqnarray}
DT_1 &=& S_{DT_1} \left[ a \left< E_1E_2 \right> \cos(\gamma) + O_1 \right]= S_{DT_1} \left[U \cos(\phi) - V \sin(\phi) + O_1^{\prime} \right] \nonumber \\
DT_2 &=& S_{DT_2} \left[ a \left< E_1E_2 \right> \sin (\gamma) + O_2 \right] = S_{DT_2} \left[ U \sin(\phi) + V \cos(\phi) + O_2^{\prime} \right] \label{eq:DTi}
\end{eqnarray}
\par\noindent are linear combinations of the Stokes
Parameters $U$ and $V$ ($C$--mode)of the polarized component of the incoming radiation.
\par\noindent In the above equations
$\gamma = \theta + \phi$ is the sum of the phase difference $\phi$
introduced by the instrument and the intrinsic phase difference $\theta$
between ${\bf E}_1$ and ${\bf E}_2$,
$S_{TPi}$ and $S_{DTi}$ are gain/conversion factors, $O_i$
are post--processing offsets and offsets produced by circuit asymmetries
and gain differences, not
completely cancelled by phase modulation and synchronous detection
(see \cite{spiga} and \cite{spie}).
% (see Section \ref{corrout}).
\par When a 90$^{\circ}$ iris polarizer is inserted between horn and orthomode
transducer the antenna splits the signal in two components
circularly polarized in opposite direction and $DT_1$ and $DT_2$
become linear combinations of $U$ and
$Q$ ($L$--mode of operation).
\par After amplification, time
integration ($\tau=6 s$) and analog to digital conversion (adc), $TP_1$,
$TP_2$, $DT_1$ and $DT_2$ are sampled three times in a $\tau$ and stored.
Each record is made of $TP_1$, $TP_2$, $DT_1$, $DT_2$,
Giulian day, UT time,
antenna pointing direction, environmental and housekeeping data.
\par MIPOL antenna and receiver are attached to a mechanical mount
which allows to move the beam along the meridian
and is driven by the same computer which stores the data.
\par MIPOL conceived at the beginning of
the '90s to check the nature of the CMB anisotropies at large angular
scales just detected by COBE-DMR \cite{cobe}, was prepared for
observation from Antarctica,
where prototypes were tested in 1994 at Terra Nova Bay \cite{BTN}
and in 1998 at Dome C - Concordia Station \cite{domec}.
\par Because in the following years we did not have the opportunity for a new
observation campaign from Antarctica, in 2002 we decided to install
MIPOL on the Italian Alps
at the Testa Grigia Observatory (lat=45.93 N,
long=7.7 E, 3480 m asl). Here it was used as a test system of
new polarization radiometers. Compared to
more recently built ground and space experiments
(e.g. PLANCK \footnote{http://www.rssd.esa.int/index.php?project=planck},
see \cite{adepar})
it is no longer competitive with the exception of its
capability of studying circular polarization.
So we decided to exploit MIPOL $C-$mode and used
data collected in 2009-2010
for improving the current upper limits of
the CMB circular polarization.

\section{Observations} \label{obs}
\par Between Nov. 10th and Dec. 9th, 2009 we set MIPOL in $C-$mode,
$7^{\circ}$ beam, and performed drift scans of the sky while the beam
moved back and forth along the meridian, at a constant pace
between $\delta_{ini}=41.1 \pm 0.1$
and $\delta_{fin}=15.8 \pm 0.1$ in 300 s, then returned to
$\delta_{ini}$ in 18 s. After a 48 s stop the cycle
started again. Every 30 minutes
the data stored on the PC hard disk were transferred via E-mail (smtp)
to our Laboratory. Except for weekly checks the system run unattended.
Weather conditions (cloud coverage, snow, wind), collected de visu when on site, or through
a webcam when in Milano, were manually recorded on the log book.
% *****************************
\begin{figure}[t]
\begin{center}
\includegraphics[angle=0,scale=.30]{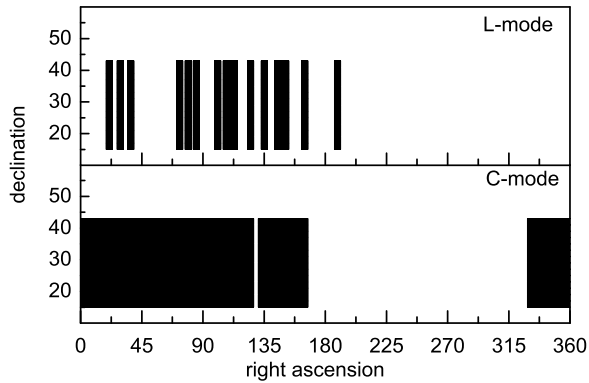}
\caption{{\it Bottom panel}: Regions of sky observed in $C-$mode;
{\it Top panel}: Regions of sky observed in $L-$mode.}
\label{fi11}
\end{center}
\end{figure}
% ****************************
\par Between Dec. 17th, 2009 and Jan. 20th, 2010 a small number
of similar drift scans were made with MIPOL in $L$--mode. Insufficient to
detect CMB linear polarization they were intended to verify MIPOL
performance and to monitor the Testa Grigia environment conditions.
Fig.\ref{fi11} shows the regions of sky covered by MIPOL during
our campaign.

\subsection{Calibration and Tests}
\label{calib}

The gain/conversion factors from digital units (adu)
to temperature (K) (see eqs.(\ref{eq:TPi}) and (\ref{eq:DTi}))
were:
\par  i) measured for $TP_1$ and $TP_2$, coupling the antenna
to an artificial blackbody source made of ECCOSORB$^{\circledR}$ AN-72\footnote{http://www.eccosorb.com/}
set at different temperatures $T_{bb}$ ranging between ambient temperature and liquid Nitrogren temperature.
\par ii) calculated for $DT_1$ and $DT_2$,
propagating the measured total power values through the phase discriminator
components whose gains and attenuations were carefully measured in
laboratory.
Offsets $O_i$ (see eqs.(\ref{eq:DTi})) can be obtained plotting $DT_i$ vs $T_{bb}$. Different samples of data
collected by MIPOL give values distributed around the average values shown in Table \ref{Ta2}. Their values are related to the un-equalized electrical offset cancellation, while their dispersions reflect the very different environmental conditions occurred during the full data taking. A more accurate evaluation of the offsets has been performed using the fitting procedure described in Sec. \ref{corrout}, on the subsample used for the subsequent analysis. Results obtained on the subsample by the two methods coincide within the error bars. Accurate evaluation of the offset values are of importance only for measurements of the monopole term. But measuring it would require precise levels set by absolute sources of polarized radiation we do not have. For this reason in the following we will ignore the monopole term. On the other hand the residual fluctuations of the offset levels, which remain after the complete analysis described in Sec. \ref{corrout}, affects the accuracy of the determination we got on the Stokes parameters at the several angular scales.

% *****************************************
\begin{table}
\begin{center}
\begin{tabular}{|cccc|}
\hline
&$C$--mode&$L$--mode& \\
\hline
Total Power &&&\\
$S_{TP_1}$&$1165.8 \pm 20.0$&$1220.3 \pm 20.8$& adu/K \\%& ~Tot.Pow. Out. 1\\
$S_{TP_2}$&$1474.0 \pm 25.4$&$1457.8 \pm 23.9$& adu/K \\%&Tot.Pow. Out. 2\\
$\Delta G_{TP1}/(\Delta t ~G_{TP1})$& $8.0~10^{-9}$ & $4.3~10^{-9}$& s$^{-1}$ \\
$\Delta G_{TP2}/(\Delta t ~G_{TP2})$& $6.4~10^{-9}$ & $4.4~10^{-9}$& s$^{-1}$ \\
\hline Correlator&&&\\
$S_{DT_1}$&$(2.9 \pm 0.9) \cdot 10^{5}$&$(3.0 \pm 0.9) \cdot 10^{5}$&adu/K\\%&Corr. Out. 1
$S_{DT_2}$&$(2.9 \pm 0.9) \cdot 10^{5}$&$(3.0 \pm 0.9) \cdot 10^{5}$&adu/K\\%& Corr. Out. 2
$O_1$&$-(0.36 \pm 0.04) \cdot 10^{5}$&$~~(1.85 \pm 0.07) \cdot 10^{5}$& adu\\%&offset 1\\
$O_2$&$-(1.79 \pm 0.08) \cdot 10^{5}$&$-(1.06 \pm 0.06) \cdot 10^{5}$ & adu\\ %&offset 2
$\Delta G_{DT1}/(\Delta t ~G_{DT1})$& $4.3~10^{-9}$ & $4.0~10^{-9}$& s$^{-1}$ \\
$\Delta G_{DT2}/(\Delta t ~G_{DT2})$& $4.4~10^{-9}$ & $8.5~10^{-9}$& s$^{-1}$ \\
\hline
\end{tabular}
\caption{MIPOL sensitivities, gain stabilities and correlator offsets.}
\label{Ta2}
\end{center}\end{table}
% **********************************
\begin{figure}[t]
\begin{center}
\includegraphics[angle=0,scale=.40]{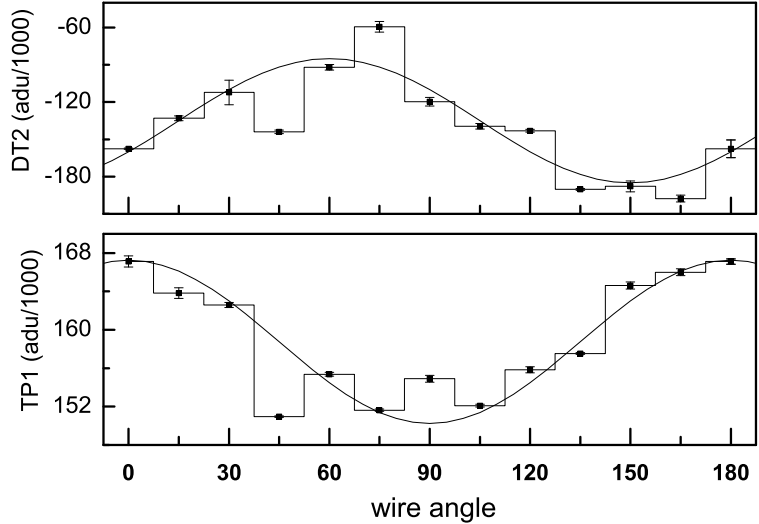}
\caption{Modulation of MIPOL outputs produced
by a rotating grid (see text)  {\it Bottom panel}: Total Power $TP_1$, $C$--mode.
 {\it Top panel}: Correlator $DT_2$, $C$--mode. Histograms: observed modulation, with statistics error bars; 
smooth curve: trend of the expected ideal modulation (see eq. (\ref{Tgrid})).}
\label{fi1}
\end{center}
\end{figure}
% ***********************************

\par Whenever necessary MIPOL behavior was checked using:
\par i) a $(5 \times 5)$ cm$^2$ flat grid of equally
spaced parallel wires (0.3 mm diameter, spaced
0.6 mm $<< \lambda$), which can be installed in the horn far field,
parallel to the horn mouth, with its normal axis coincident with
the horn axis. Crossed by
the sky radiation, it injects in the horn a linearly polarized
wave of temperature
\begin{eqnarray}
T_p \simeq (\Omega_g/\Omega_h) [T_{sky} \sin^2 (\theta_w) + T_{back} \cos^2 (\theta_w)]
\label{Tgrid}
\end{eqnarray}
where $T_{sky}$ is the sky temperature, $\Omega_g$ is the solid angle of the grid seen
from the horn center of phase and $\Omega_h$ the solid angle of
the horn beam ($\Omega_g/\Omega_h \sim 0.19$), $T_{back}$ is the noise back-reflected by the grid into the horn and $\theta_w$ is the angle respect to the wires direction. Rotating the grid around its vertical axis we can modulate
both total power (single mode and single polarized) channels and correlator outputs in a way which in principle is
a simple sine law, proportional to the smooth curves shown in
fig. \ref{fi1}. However grid back-reflection of the
noise radiated by the horn and by the surrounding shield and
horn mismatches produced by the grid itself and by grid supports
make the effective modulations (histograms in fig. \ref{fi1})
more involved than that suggested by equation \ref{Tgrid}. In addition $T_{back}$ is not easy to be estimated and, due to the not circularly symmetric grid support structure, should be dependent on $\theta_w$. The grid signal
therefore cannot be used for accurate calibrations. Grid
modulation turns out however very useful for checking
that MIPOL remained sensitive to polarization: tiny quantities of dust or water vapor
(produced by melting snow and ice needles) in the horn throat are
in fact sufficient to make MIPOL deaf and completely cancel
the dependence of $TP_i$ and $DT_i$ on the rotation angle.
The grid signal depends on the environmental conditions, but it is well reproduced when similar conditions occur.
\par ii) a solid state noise generator\footnote{Hewlett-Packard HP R347B}
which can be used to inject via directional couplers
similar signals in both receiver channels ($T_1^{NG}=( 3.17 \pm 0.04)$ K and
$T_2^{NG}=( 4.29 \pm 0.04)$ K in $C$--mode, $T_1^{NG}=( 2.87 \pm 0.02)$ K and
$T_2^{NG}=( 3.42 \pm 0.03)$ K in $L$--mode). These signals, produced
four times a day for 15 minutes, have been used to work out the gain
stabilities of Mipol channels shown in Table \ref{Ta2}.

\section{Data Reduction} \label{dar}

Only data collected at nighttime (i.e. between half an hour after
sunset and half an hour before sunrise) have been analyzed. We then
eliminated records which: i) contained anomalous housekeeping data, ii) showed
odd values or odd variations of the receiver outputs, iii) were associated to bad weather
conditions or to incomplete zenith scans.
Here and in the following rejecting a value of $DT_1$ or $DT_2$ or $TP_1$ or $TP_2$
automatically causes
rejection of the complete data record therefore of $DT_1$ and $DT_2$ and
$TP_1$ and $TP_2$ and of all the records associated to the same zenith scan.
\par Right ascension $\alpha$  and declination $\delta$
of the beam axis were then calculated and added to each record. Records
associated to tests and calibrations were separated
and used to work out the system sensitivities $S_{TP_i}$ and $S_{DT_i}$.
Finally data were compressed in declination ($\delta$) bins of $2^{\circ}$.

\subsection{Total Power outputs}
\label{totpow}
MIPOL does not include absolute references of temperature,
therefore $TP_1$ and $TP_2$ have been used for monitoring
environment conditions and MIPOL behavior, not for
measurements of the CMB absolute temperature or anisotropy.
\par We expect:

\begin{eqnarray}
{TP_i \over  S_{TP_i}} (\alpha,\delta,z) = T_i^{sky} (\alpha,\delta) + T_i^{atm}(z) + T_i^{gr}(z) + T_i^{rx}
\label{Totalp1}
\end{eqnarray}
\par\noindent where
\begin{eqnarray}
 T_i^{sky} (\alpha,\delta) = T_i^{CMB} + T_i^{gal}(\alpha,\delta) + T_i^{ex}
\label{Totalp2}
\end{eqnarray}
\par\noindent Here, $z$ is the beam zenith angle, $T_i^{CMB}, T_i^{gal}(\alpha,\delta)$ and $T_i^{ex}$ the brightness
temperature of CMB, Galactic emission and blend of unresolved extragalactic sources, respectively. $T_i^{atm}(z)$ and $T_i^{rx}$ are
the atmosheric signal and the receiver noise while
$T_i^{gr}(z) = T_{i,min}^{gr}+ \Delta T_i^{gr}(z)$ is the contribution of
ground and
other undesired emission from the environment, which overcome the antenna
ground screen.
Because of obstacles northward of MIPOL axis, $T_i^{gr}(z)$ was minimum
when z is close to 0, not exactly at $z=0$.
At 33 GHz and MIPOL angular resolution
we expect $T^{gal}\lesssim 4$ mK \cite{tris1,tris3},
$T^{ex} \approx 15 \mu$K \cite{trisger} and
$\Delta T_i^{gr}(z) \ll  T_i^{atm}(z)$ so to first approximation:
\begin{equation}
{TP_i \over S_{TP_i}} \simeq T_i^{CMB} + T_i^{atm}(z) +T_i^{sys}
\end{equation}
where $T_i^{sys}= T_i^{rx} + T_{i,min}^{gr}$ is the system noise.

Total power analysis was carried out for the full samples
of $C-$ and $L-$mode
data, and then repeated for the subsample of $C-$mode data on which
the correlator analysis was performed (see Section \ref{corrout} for details).
\par
\begin{figure}[hp]
\begin{center}
\includegraphics[width=5cm,angle=-90]{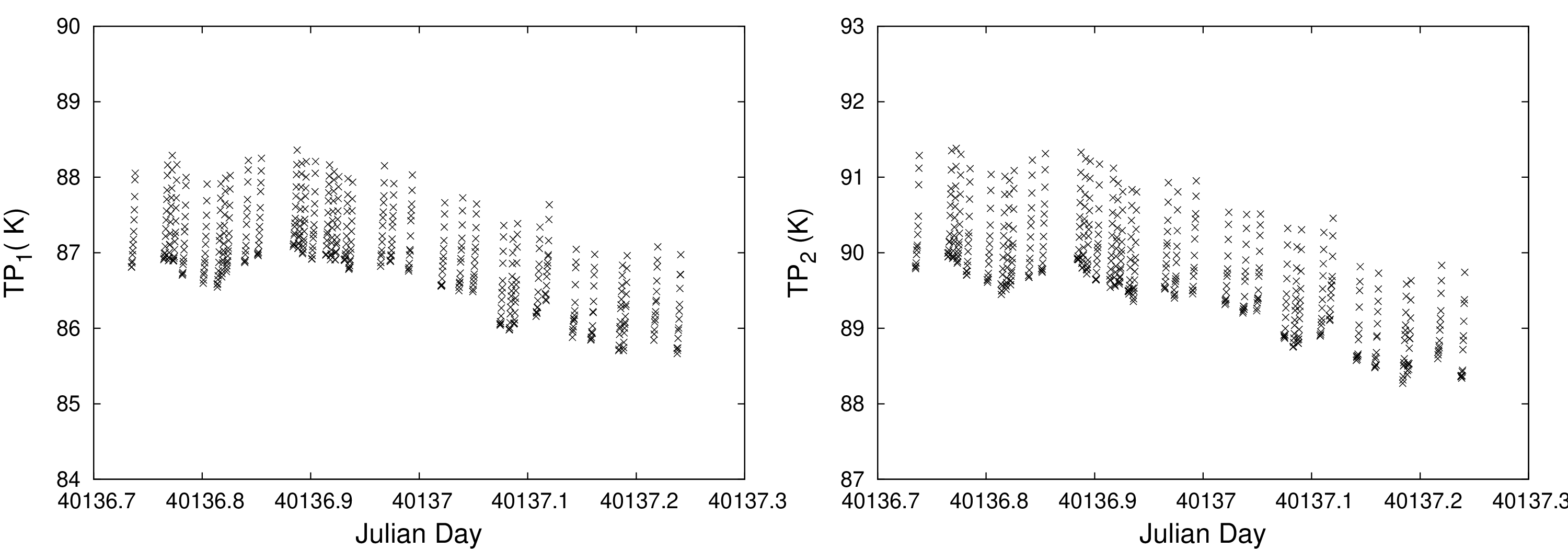}
\includegraphics[width=5cm,angle=-90]{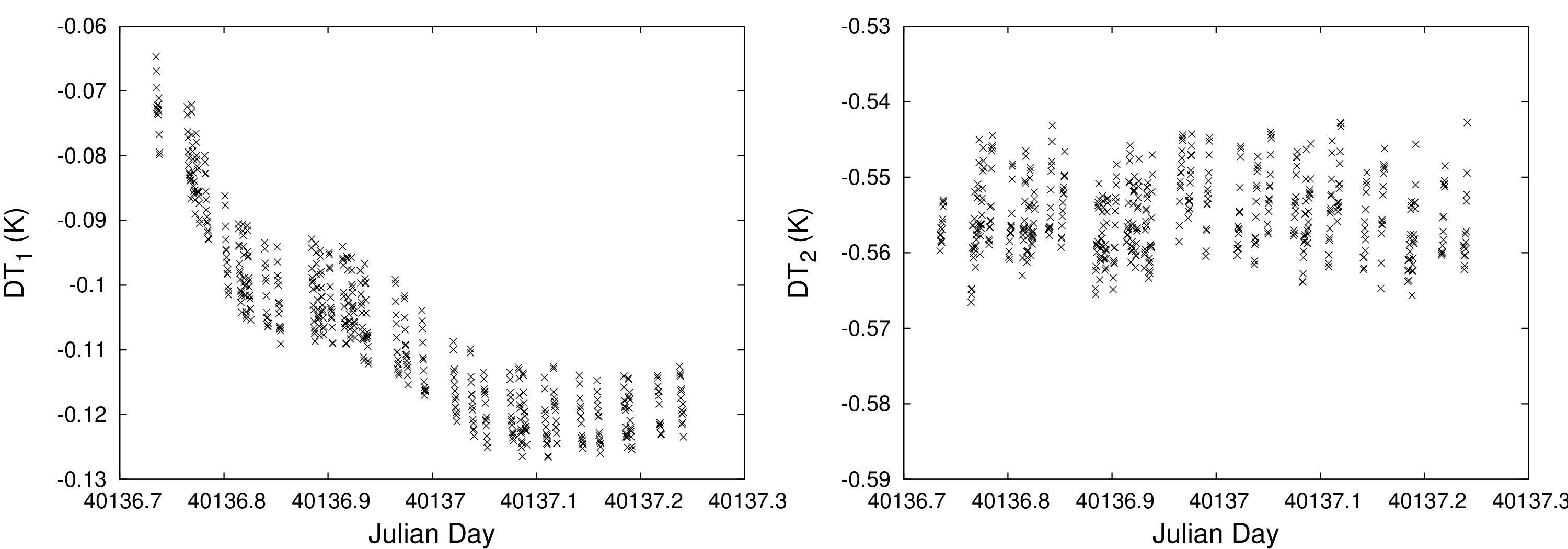}
\caption{$TP_i$ and $DT_i$ TOD subsample (see text)  .
}\label{f1}
\end{center}
\end{figure}

 The Time Ordered Data (TOD) restricted to the subsample are
displayed in Fig. \ref{f1} for both $TP_i$ and $DT_i$. The same set of data is then used in all
the subsequent plots where, at difference from Fig. \ref{f1},
the mean values of the zenith scans have been equalized to their average value.
\par For each sample, the atmospheric noise temperature $T_i^{atm}$ has been
extracted from the zenith scans fitting them with a secant law
$T_i^{atm}(z)=T_i^{atm}f(z)$  where $f(z)=\sec(z)\ast \cal {G}$ is the
convolution of $\sec (z)$ over the antenna beam $\cal {G}$ (which is assumed
to be Gaussian).

% ********************************
% ********************************

\begin{figure}[]
\begin{center}
\includegraphics[width=5cm,angle=-90]{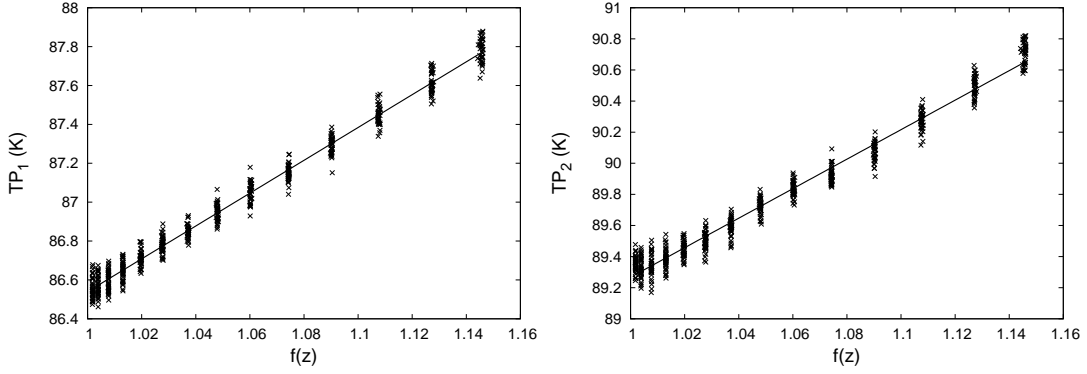}
\caption{Total Power zenith scan profiles $TP_i$.
Data are compressed in declination bins of $2^{\circ}$ and plotted versus
$f(z)=\sec(z) \ast \cal {G}$, the convolution of $sec(z)$ over the antenna beam
$\cal {G}$. Data are well fitted by a secant law (solid lines).
}\label{f2}
\end{center}
\end{figure}

The resulting values are consistent with model expectations \cite{ajello} and
summarized in Table \ref{t1}. $C-$mode fits are shown in Fig. \ref{f2}.
\par Differences between the atmospheric temperatures
extracted from the full samples of data obtained while MIPOL was in $C-$ and $L-$mode
are consistent with variability of weather conditions: frequent perturbations
and higher nighttime temperatures ($<T_{env}> = -9.3 \pm 4.0$ C)
in November - beginning December 2009, clear sky, stable
weather conditions and colder
nights ($<T_{env}> = -17.2 \pm 5.6$ C), at the end of December 2009
and in January 2010.

More noticeable are the differences between the $C-$mode results
obtained from the full sample and the subsample.
They reflect the large variability of observing and
system conditions during the entire period of observations compared
to the stable conditions, which produced uniform sky coverage, during the night
when the entire subsample was collected (see next Section).

\par Subtractions of the atmospheric signals from $TP_i$ zenith scans
leaves a residual $z$ dependence
produced by the increasing fraction
of ground emission which overcomes the ground screen as the antenna
moves along the meridian toward the horizon.
After subtraction of atmospheric signal and ground excess $\Delta T_i^{gr}(z)$,
$TP_1(\alpha,\delta)$ and $TP_2(\alpha,\delta)$ become $z$ independent.
Removing  $T^{CMB}=2.018$ K, the brightness temperature of the $CMB$ at 33 GHz,
gives the system noise $T_i^{sys}$ (see Table \ref{t1}).
% ******************************************
\begin{table}
\begin{center}
\begin{tabular}{|ccccc|}
\hline
&  $L-$mode & $C-$mode & $C-$mode&  \\
&  full sample & full sample & subsample& \\
\hline
 $T_1^{atm}$   &$9.55 \pm 0.21$ & $7.91 \pm 0.04$&$8.45\pm 0.04$ & K \\
  $T_2^{atm}$  & $7.25 \pm 0.26$ &$8.70 \pm 0.05$ &$9.49 \pm 0.06$& K \\
 $T_1^{sys}$  &$ 83.6 \pm 1.4$ &$82.9 \pm 1.4$  &$75.36\pm 0.05$& K \\
  $T_2^{sys}$ &$ 89.7 \pm 1.5$ &$84.1 \pm 1.5$&$ 76.05 \pm 0.07$& K \\
\hline
\end{tabular}
\caption{Atmospheric and system temperatures from Total Power measurements.
Results are reported for the full sample of data (both for $C-$mode and
$L-$mode) and the $C-$mode subsample described in the text.}
\label{t1}
\end{center}
\end{table}
% ************************************************

\subsection{Correlator outputs}
\label{corrout}
Because of the limited quantity of $L-$mode data and
our interest for circular polarization,
here and in the following only $C-$mode data have been used.
\par
Shapes of the correlator zenith scans are poorly defined and
depend on weather conditions. Moreover their base levels
vary day by day because of: i) slow variations
of gain and system noise (not evident on $TP_i$ profiles
because of the very different sensitivities of correlator
and total power (see Table \ref{Ta2})); ii) polarization by reflection of
ground contribution (see section \ref{totpow}) and dependence of
the ground reflectivity on the humidity.
\par
The set of $C-$ mode correlator data can be divided
in two groups: i) a subsample of data collected
in the night between Julian days 40136 and 40137, when the
observing conditions were particularly good. Characterized by
$rms$ fluctuations of $DT_1$ and $DT_2$ equal to
$\sigma_1 \simeq 0.7 \ mK$ and $\sigma_2 \simeq 0.9 \ mK$ respectively,
the data of this subsample fill uniformly the region of sky observed by
MIPOL, show a uniform distribution respect to elevation and time,
and represent $1/3$ of the complete sample of $C-mode$ data; ii)
the remaining data characterized by $\sigma_1 \simeq 1.8 \ mK$
and $\sigma_2 \simeq 1.9 \ mK$. They are definitely more noisy,
their distribution on the sky is irregular and were collected when the observing
conditions were definitely worse. Combining the two sets of data
the statistics increases but fluctuations are
$\sigma_1 \simeq 1.5 \ mK$ and $\sigma_2 \simeq 1.6 \ mK$,
definitely worse than the subsample values.
We decided therefore to concentrate our analysis using the smaller but
cleaner subsample only.
\par Let us rewrite eq. (\ref{eq:DTi}) emphasizing the contribution of the different
sources:
\begin{eqnarray}
\nonumber
{DTi \over S_{DTi}}(z)
& = &  {1\over S_{DTi}}\sum_X DT_i^X  \nonumber \\
& = & a\sum_X  \left<E_1^XE_2^X\right> h_i(\gamma_X) +O_i
\label{diff1}
\end{eqnarray}
\par\noindent where we made use of the correlation properties of radiation
($\left< E_1^XE_2^Y \right> =0$ for $X \neq Y$) and set
$h_1(\gamma_X)= \cos(\gamma_X)$, $h_2(\gamma_X)=\sin(\gamma_X$) and
$\gamma_X=\theta_X+\phi$
\par Marking the $z$ dependence we can write:
\begin{eqnarray}\label{fg}
{DTi \over S_{DTi}}(z)
& = &  a[\left< E_1^{atm}E_2^{atm} \right>_0 f^c(z) h_i(\gamma_{atm}) +  \left<E_1^{gr}E_2^{gr}\right>_0 g^c(z)h_i(\gamma_{gr})] + O_i
\label{diff2}
\end{eqnarray}
\par\noindent where: i) $\left< ~~ \right>_0$ marks correlated signals at $z=0$; ii)
no receiver noise correlation term is present because
the noises in channel 1 and 2 are generated independently; iii) the
sky signal, dominated by $CMB$, has a negligible $z$ dependence; iv) the
$z$ dependences of atmospheric and ground signals are described by
$f^c(z)$ and $g^c(z)$ respectively.
% ***********************************
\begin{table} \label{t33}
\begin{center}
\begin{tabular}{|cccc|}
\hline
  $\tan(\gamma_{atm})$ & $ T_{corr}^{atm,0}$ (K) & $\tan(\gamma_{gr})$  & $ T_{corr}^{gr,0}$ (K)  \\
\hline
$-1.68\pm 0.09$ &$0.087\pm0.002$ &$-0.24\pm0.06$ &$0.006\pm0.046$ \\
\hline
\end{tabular}
\caption{Correlator phase differences $\gamma_{X}$ (see text) and estimated
atmospheric and ground polarized emissions.}
\label{t2}
\end{center}
\end{table}
% ****************************************
\par
\begin{figure}[]
\begin{center}
\includegraphics[width=5cm,angle=-90]{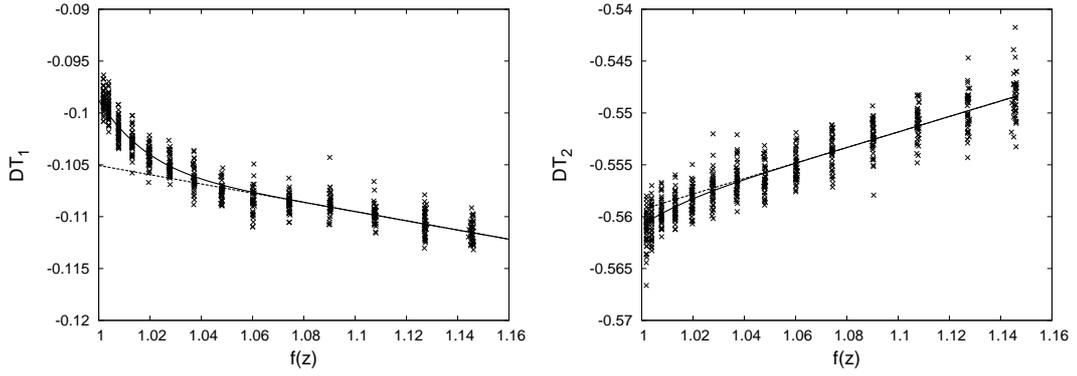}
\caption{Same as in Fig. \ref{f2} but for the correlator zenith scan profiles
$DT_i$. Deviations from a pure secant law (dashed lines) at small zenith angles
are here evident due to ground contaminations and unknown enviromental effects.
Data at small angles are well fitted by polinomyals of order $8$ in $f(z)$
(solid lines).}
\label{f5}
\end{center}
\end{figure}

 Setting $f^c(z)=\sec(z)\ast \cal {G}$, we can fit the $DT_i$ zenith
scan profiles adding a term of order 8 in $f^c(z)$,
which accounts for $g^c(z)$, (see Fig. \ref{f5}). We get
\begin{equation}
DT_i(f^c(z))  =  d_{0,i} + d_{1,i}f^ c(z) + d_{8,i} (f^c(z)-1.146)^8
\label{fitdiff}
\end{equation}
% ********************************
% ********************************
% ********************************
\begin{figure}[]
\begin{center}
\includegraphics[width=5cm,angle=-90]{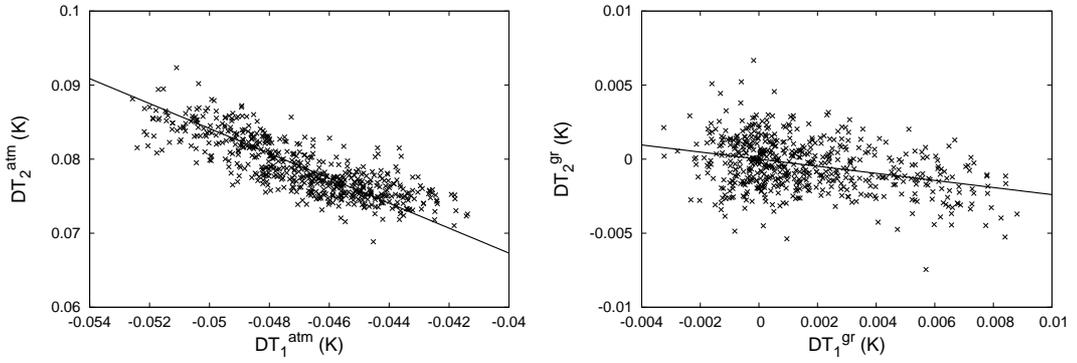}
\caption{$DT_2$ versus $DT_1$ plots for atmospheric ({\it left}) and ground
({\it right}) signals. The angular coefficients of the fits (solid lines) give
the phase differences $\gamma_{atm}$ and $\gamma_{gr}$ (see text).
}
\label{f6}
\end{center}
\end{figure}
% ********************************
\begin{figure}[]
\begin{center}
\includegraphics[width=5cm,angle=-90]{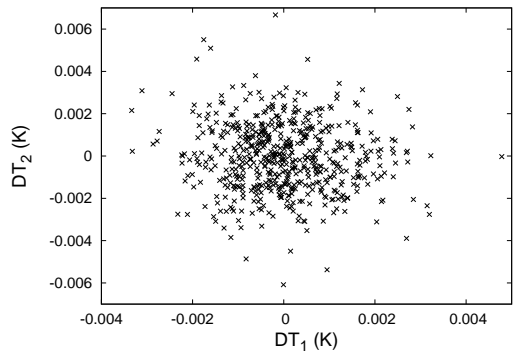}
\caption{$DT_2$ versus $DT_1$ plot after removing atmospheric and ground
emissions and offsets (see text).}
\label{f6bis}
\end{center}
\end{figure}
% ********************************
\par\noindent Comparing eqs.(\ref{fitdiff}) and (\ref{diff2}), the
coefficients $d_{1,i}$ and $d_{0,i}$ of the linear term
(see Fig. \ref{f5})
\begin{equation}
DT_i^{lin}(f^c(z))  =  d_{0,i} + d_{1,i}f^c(z)
\label{fitlin}
\end{equation}
can be identified with
$\left< E_1^{atm}E_2^{atm} \right>_0 h_i(\gamma_{atm}) $ and $O_i$ respectively.
\par\noindent The resulting values of $O_i$ {\bf ($O_1 = -(0.177 \pm 0.006) \cdot 10^{5}adu =
-(0.061 \pm 0.002) K$, $O_2 = -(1.839 \pm 0.009) \cdot 10^{5}adu = -(0.634 \pm 0.003) K $))}.
Only the monopole term, we will not consider here,
depends on these values. The fluctuation of $O_i$, after removing the residual linear dependence
with the environment temperature, described later in the present section, set the final accuracy
on the Stokes parameters at the several angular scales.
The last term in eq. (\ref{fitdiff}), which describes
deviations from the secant law, accounts for ground excess
$\left< E_1^{gr}E_2^{gr} \right>_0 g^c(z) h_i(\gamma_{gr}) $ and other
unknown contributions.
\par We can now obtain the phase differences
$\gamma_{atm}= \tan^{-1} \left({d_{1,2}/ d_{1,1}}\right)$ and
$\gamma_{gr} =\tan^{-1} \left({d_{8,2}/ d_{8,1}}\right)$: they are
the angular coefficients of the lines shown in Fig. \ref{f6}
(alternatively they can be obtained plotting $DT_2^X$ versus
$DT_1^X$). Finally we get the correlated (therefore polarized)
components of the atmospheric
$T_{corr}^{atm,0}\propto \left< E_1^{atm}E_2^{atm} \right>_0$ and ground
$T_{corr}^{gr,0} \propto \left< E_1^{gr}E_2^{gr} \right>_0$ emissions (see
the next Section for a discussion on the atmospheric signal).

After removing them and offsets, and correcting for a residual linear
dependence of the resulting time profiles of $DT_i$
at constant $z$ on the environment temperature,
no significant correlation is observed between $DT_1$ and $DT_2$
(correlation coefficient $\rho = -0.09$, see Fig \ref{f6bis})
indicating the polarized signal is completely buried in the instrumental noise.

Furthermore, the expected degree of polarization of the
atmospheric emission, if present,
must be very low so that the signal can be supposed unpolarized and
we can assume $\theta_{atm} << \phi$. It follows
$\gamma_{atm} \simeq \phi$ and $\gamma_{gr}= \theta_{gr} +\phi$.
The resulting values of $T_{corr}^{X,0}$ and $\gamma_X$ ($X=atm,gr$) are
summarized in Table \ref{t2} together with the $1-\sigma$ uncertainties.

We can therefore work out the sky Stokes parameters U and V inverting:
\begin{eqnarray}
DT_1 & \propto &  U \cos(\phi) - V \sin(\phi)  \nonumber \\
DT_2 & \propto &  U \sin(\phi) + V \cos(\phi)
\end{eqnarray}
The resulting Stokes parameters $U$ and $V$ of the sky signal
were finally arranged
in squared bins to form maps with resolution
$2^\circ, 8^\circ, 12^\circ$ and $24^\circ$.
The maps are noise dominated and do not show statistically significant features.

% ******************************************
\begin{table}
\begin{center}
\begin{tabular}{|ccccccc|}
\hline
 pixel size& pixels  & $\left<V\right>$ & $\sigma_V$ & $\sigma_V^{pix}$ & $\sigma_{\left<V\right>}^{pix}$ &\\
\hline
$2^\circ \times 2^\circ$ &&&&&&\\
unbinned data& 518 & -0.02 $\pm$ 0.06&  1.29 &&& mK\\
\hline \hline
 $8^\circ \times 8^\circ$ &64&-0.02 $\pm$ 0.08 & 0.64 & 1.14 & 0.43 & mK\\
 $12^\circ \times 12^\circ$ &32& -0.02 $\pm$ 0.07 & 0.38  & 1.27 & 0.35 & mK\\
$24^\circ \times 24^\circ$ &9&-0.02 $\pm$ 0.06 & 0.15  & 1.37 & 0.21 & mK\\
\hline
\end{tabular}
\caption{CMB Stokes Parameter V from an unbinned map, dominated by the system
noise, insensitive to  sky signal (upper section) and three maps at
different angular resolutions sensitive to sky signal (lower section).
The Table columns show: pixel size, number of pixels, mean
$\left<V\right>$, standard deviation $\sigma_V$ over the full maps,
average $rms$ per pixel $\sigma_V^{pix}$ and average
standard deviation of the mean per pixel $\sigma_{\left<V\right>}^{pix}$.}
\label{t3}
\end{center}\end{table}
% ***********************************
\par\noindent Concentrating our attention on $V$ circular polarization
Stokes parameter for each $V$ map we calculated (Table \ref{t3}): i) $V$ mean value
$\left<V\right>$
(with $1-\sigma$ uncertainty); ii) standard deviation $\sigma_V$ calculated
over the full map; iii) average value of the $rms$ per pixel $\sigma_V^{pix}$;
iv) average standard deviation of the mean per pixel $\sigma_{\left<V\right>}^{pix}$; v)
$\left<V\right>$ and $\sigma_V$ for the un-binned data.
\par Because the bins of the $2^\circ \times 2^\circ $ map are smaller than
MIPOL angular resolution, at this angular scale the sky signal is washed out and the system noise dominates. Therefore
in the following the $V$ and $\sigma$ values of the $2^\circ \times 2^\circ $ map
will be marked $V_{noise}$ and $\sigma_{noise}$. For the map at $8^\circ, 12^\circ$ and
$24^\circ$ the $V$ and $\sigma$, combinations of noise
and sky signal, will be marked $V_{obs}$ and $\sigma_{obs}$.

\section{Discussion} \label{dis}
\subsection{Polarized component of the atmospheric signal}
\label{Atmo}
\par\noindent Circular polarization of the atmospheric emission
is produced by Zeeman effect on the oxygen molecules therefore
depends on the angle $\epsilon$ between the line of sight and
the geomagnetic field line at the site of observation
\footnote{$http://omniweb.gsfc.nasa.gov/vitmo/cgm_vitmo.html$}.
Following \cite{spinelli}: i) $f_{O_2}^c = \sec(z) \cos(\epsilon)$;
ii) at the Testa Grigia Observatory (lat=45.93 N, long=7.7 E)
the Earth magnetic field line is approximately on the meridian plane,
points northward and the angle between field line and zenith direction is
$\epsilon_0 \simeq$ 151.9 deg; iii) at MIPOL
operating frequency (33 GHz) we can expect
$V_{O_2}^{33GHz} \sim 50-70 \ \mu K$ at $z=0$ with an increase
($\Delta V \sim 15-25 \ \mu K$) looking  $z=30$ deg southward.
So we can write $\epsilon \simeq z + \epsilon_0$ and
$f_{O_2}^c (z) = \sec(z) \cos(z + \epsilon_0)$.
\par\noindent Fitting our zenith scan profiles of
$DT_1$ and $DT_2$ assuming  $f^c(z) \simeq f_{O_2}^c(z)$
we obtain results compatible with the results previously obtained
assuming $f^c(z)=\sec(z)$.
Differences between the two fitting functions are in fact small.
The differences become important close to $z=0$,
where the increase of the ground contribution
produced by northward obstacles makes this analysis unreliable.
Last but not least the signal we obtain
is too large and cannot be associated to polarized
emission of the atmospheric oxygen.
We are rather lead to ascribe the signal correlated to the scans to $I \rightarrow (V,U)$
contamination of the correlator outputs by the
much stronger unpolarized atmospheric signal measured in total power.
\par\noindent Therefore $T_{corr}^{atm,0}$ is an upper limit to the circularly
polarized emission of the atmospheric oxygen and we can write
$V_{O_2}^{33GHz}(z=0) < 87$ mK and quantify the $I \rightarrow (V,U)$
contamination by
\begin{equation}
\Gamma(I \rightarrow V,U) = \frac{T_{corr}^{atm,0}}{\sqrt{T_1^{atm,0}T_2^{atm,0}}} = 9.7 \times 10^{-3}
\label{leakage}
\end{equation}

\subsection{Sky signal}
\label{residuals}
\par
The monopole term of $V$ and the very large scale
fluctuations approximately constant over the map size are hidden in the
large instrumental offsets and can not be discriminated with our measurement
apparatus. Removing offsets then makes the average values of $V$, in our maps
consistent with zero at the angular scales we considered
(see Table \ref{t3}).

Upper limits on the degree of the CMB circular polarization are obtained with 
three different methods (see below) which rely on the assumption 
that data are drawn from a Gaussian distribution.
In order to check the validity of this assumption, we 
perform a $\chi^2$ test for the distributions of the measured $V$ values
at angular scales of $2^\circ$, $8^\circ$ and $12^\circ$. Because of the 
small number of independent pixels, 
$\chi^2$ test is meaningless at $24^\circ$ resolution.
For each scale, we assume a Gaussian model with mean $\mu$ and 
standard deviation $\sigma$ estimated from:

i) data, assuming $\mu=\left< V \right>$ and $\sigma = \sigma_V$ (see 
Table \ref{t3}) or, alternatively, best--fitting the observed distibutions; 

ii) Monte Carlo simulations. We generate $10000$ random Gaussian 
realizations of our $2^\circ \times 2^\circ$ map (same number of data, mean and
variance), and then bin at $8^\circ$ and $12^\circ$. Model parameters 
$\mu$ and $\sigma$ are evaluated from the distributions of the simulated data
at the angular scales considered. 

In Table \ref{tchi2} we list the number of degree of freedom ${\it f}$, 
$\chi^2$, and the corresponding probability $P$ of the assumed Gaussian model. 
At $2^\circ$, the distribution of the data is Gaussian to a very 
good approximation.
Al larger angular scales, we can claim that data are consistent, 
at $95 \%$ C.L., with the hypothesis of Gaussianity.

\par All the circularly polarized signals, both those of
astrophysical origin (Galactic synchrotron,
blend of unresolved extragalactic sources, SZ effect), and
those of cosmological origin, associated to the CMB, are expected at
the $\mu K$ level or below.
Therefore all the sky signals at $8^\circ$, $12^\circ$ and $24^\circ$ are
completely buried in the MIPOL noise and the $rms$ per
pixel, $\sigma_V^{pix}$, is consistent with the fluctuations
$\sigma_V$ of the unbinned ($2^\circ \times 2^\circ$) data. 
For each map we also expect $\sigma_V \simeq \sigma_{\left< V \right>}^{pix}$.
Slight deviations found at $8^\circ$ and $24^\circ$ scales
reflect the limited statistics of those maps
(few measurements per pixel at $8^\circ$ resolution,
small number of pixels at $24^\circ$).
% **************************************
\begin{table}
\begin{center}
\begin{tabular}{|c|c|cc|cc|cc|}
\hline
\multirow{2}{*}{pixel size}&\multirow{2}{*}{\it f} & \multicolumn{2}{c|}{$\mu=\left< V \right>$ $~\sigma = \sigma_V$}&\multicolumn{2}{c|}{best--fit}&\multicolumn{2}{c|}{simulations}\\
%\hline
 & &$\chi^2$&$P$&$\chi^2$&$P$&$\chi^2$&$P$\\
\hline
 $8^\circ \times 8^\circ$ &$5$&$6.04$  & $0.70$ &3.90 &$0.44$ &$9.95$&$0.92$  \\
 $12^\circ \times 12^\circ$&$1$& $1.96$ &$0.76$&$0.79$ &$0.55$&$2.81$ &$0.83$   \\
\hline \hline
 $2^\circ \times 2^\circ$&$55$& \multicolumn{6}{c|}{$\chi^2 = 46.60 ~~~ P=0.20$}\\
\hline
\end{tabular}
\caption{$\chi^2$ test for assessing the Gaussianity of the data. 
We list the number of degree of freedom ${\it f}$, 
$\chi^2$, and the corresponding probability $P$ of the assumed Gaussian model.
See the text for details.}
\label{tchi2}
\end{center}\end{table}
% **************************************

\par For each of the above components of the sky signal and their sum
(no component stands up above the others) we use three different approaches
in order to set 95$\%$ upper limits at the angular scales of $8^\circ$,
$12^\circ$ and $24^\circ$: 
\par 
{\bf Method I} - We can write:
\begin{equation}
\sigma^2_{obs}(\theta) = \sigma^2_{sky}(\theta) + \sigma^2_{noise}(2^\circ \times 2^\circ)/(\theta/2^\circ)^2
\end{equation}
\par\noindent where $\sigma_{noise}(2^\circ \times 2^\circ) = 1.29$
, $\sigma_{obs}=\sigma_V$ (see Table \ref{t3}), and get
$\sigma_{sky}(\theta)$. The resulting upper limits (at 95\% CL) to the degree of
circular polarization of CMB, $\Pi_V^{CMB}= w/T^{CMB}$, ($w$ is the width of the probability distribution
at $95\%$ and in this case $w=2\sigma_{sky}$) are shown
in Table \ref{t12}.
\par
{\bf Method II} - A Bayesian approach \cite{partridge} allows a different estimate.
We can write
\begin{equation}
P(\sigma_{sky}|\{V_i\}) = {\cal N} ~p(\sigma_{sky}) ~P(\{V_i\}|\sigma_{sky})
\end{equation}
\par\noindent where $P(\sigma_{sky}|\{V_i\})$ is the probability that $\sigma_{sky}$
is consistent with the observed distribution $\{V_i\}=V_{obs}$,
\begin{equation}
P({V_i}|\sigma_{sky}) = \prod_i {1\over \left[2 \pi(\sigma_i^2 + \sigma_{sky}^2)\right]^{1\over2}} ~e^{-{1\over 2}{V_i^2 \over \sigma_i^2 + \sigma_{sky}^2}}
\end{equation}
\par\noindent is the likelihood distribution of the measurements ($\sigma_i$ is the standard deviation of the $i^{th}$ measurement),
$p(\sigma_{sky})$ the probability density of $\sigma_{sky}$ assumed flat (uniform
prior) and ${\cal N}$ a normalization constant. The values of $\sigma_{sky}$ above which
lay less than $5\%$ of the probability curve, reported in Table \ref{t12}, represent the upper limits (at $95\%$ CL)
for the circular polarization degree.

% *****************************
\begin{figure}[t]
\begin{center}
\includegraphics[angle=-90,scale=.30]{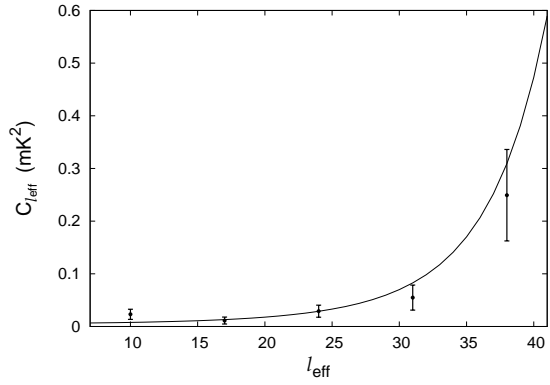}
\caption{Binned angular power spectrum $C_{{\it l}_{eff}}$ from data
(signal plus noise) compared to the expected power spectrum (solid line)
for pure Gaussian noise.}
\label{spettron}
\end{center}
\end{figure}
% ****************************
\begin{figure}[t]
\begin{center}
\includegraphics[angle=-90,scale=.30]{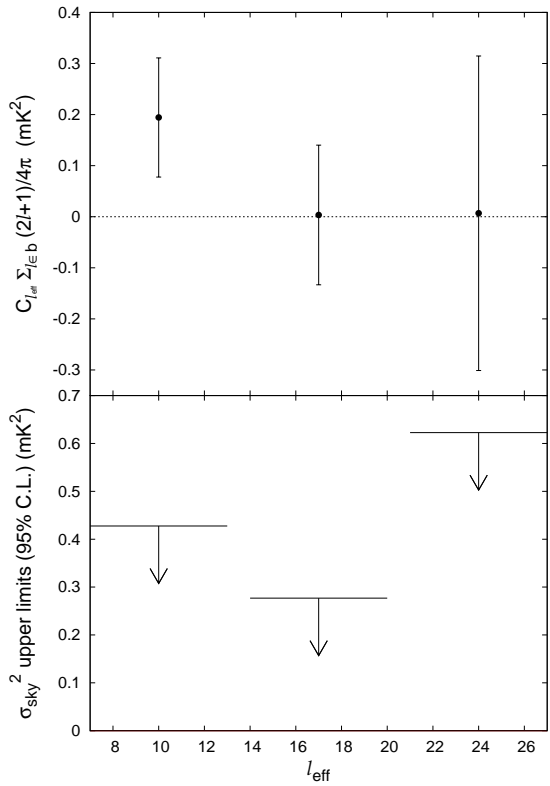}
\caption{
{\it Top panel}: Angular power spectrum of $V$-mode polarization fluctuations;
{\it Bottom panel}: $95 \%$ C.L. upper limits.}
\label{spettro}
\end{center}
\end{figure}
% ****************************
%****************************
\begin{table}
\begin{center}
\begin{tabular}{|ccccc|}
\hline
method&& I && II \\
\hline
pixel size  &&& $\Pi_V^{CMB}$ &\\
\hline
 $8^\circ \times 8^\circ$ &&$5.0\cdot 10^{-4}$&&$4.0\cdot 10^{-4}$\\
 $12^\circ \times 12^\circ$ &&$2.7\cdot 10^{-4}$&&$3.2\cdot 10^{-4}$\\
 $24^\circ \times 24^\circ$ &&$0.7\cdot 10^{-4}$&&$2.1\cdot 10^{-4}$\\
\hline
\end{tabular}
\caption{Upper limits to the degree of circular polarization
of the CMB calculated by classical methods and bayesian methods (see the text).}
\label{t12}
\end{center}\end{table}
%****************************
\begin{table}
\begin{center}
\begin{tabular}{|ccccc|}
\hline
\multicolumn{5}{|c|}{method III - spherical harmonics analysis}\\
\hline
$\Delta {\it l}$  && $<\theta>$  && $\Pi_V^{CMB}$ \\
\hline
  $7-13$ & & $18.0^{\circ}$ & &$2.7 \cdot 10^{-4}$\\
 $14-20$ & & $10.6^{\circ}$ & &$2.4 \cdot 10^{-4}$\\
 $21-27$ & &  $7.5^{\circ}$ & &$4.3 \cdot 10^{-4}$\\
\hline
\end{tabular}
\caption{Upper limits to the degree of circular polarization
of the CMB calculated by spherical harmonics analysis methods (see the text).}
\label{t13}
\end{center}\end{table}
% **************************************
{\bf Method III - Spherical harmonics analysis} - Finally, we apply the maximum likelihood method described in \cite{bond}
to estimate the angular power spectrum $C_{\it l}$ of $V$--mode polarization.
The method is a trivial application of Bayes' theorem and assume that
fluctuations in both sky signal and experimantal noise are Gaussian.
$C_{\it l}$ 's are estimated with a quadratic estimator which can be
derived from a Gaussian approximation to the likelihood function:
\begin{equation}
{\cal L}(C_{\it l})= P({V_i}|C_{\it l})={1 \over \left[2\pi^{N_{pix}} {\det
C}\right]^{1/2}} ~e^{-{1\over 2}V_i C_{ij}^{-1} V_j}
\end{equation}
where $N_{pix}$ is the number of pixels in the map and the total covariance
matrix $C_{ij}=S_{ij}(C_{\it l}) + N_{ij}$ is the sum of the theoretical signal
covariance matrix $S_{ij}$, which depends on the pameters to be estimated,
and the instrumental noise covariance matrix $N_{ij}$ estimated
from data (assuming uncorrelated and Gaussian noise).

The Gaussian approximation is equivalent to truncating the Taylor series
expansion of $\ln {\cal L} (C_{\it l}+ \delta C_{\it l})$ to second order term.
This allows to solve iteratively for $C_{\it l}$'s that maximize
${\cal L} (C_{\it l})$ starting from an initial guess $C_{\it l}^0$:
$$
C_{\it l}^{i+1}=C_{\it l}^{i}+\delta C_{\it l}
$$
with
$$
\delta C_{\it l}={1\over 2} \sum_{\it l'} F_{\it ll'}^{-1} {\partial {\ln \cal L}\over
\partial {C_{\it l'}}}
$$
where $F_{\it ll'}$ is the $C_{\it l}$'s Fisher matrix.

Given the limited extent of our map, features in the power spectrum will be
smeared out on scale smaller than $l\sim \pi/ \theta$, ($\theta$ is the size of
the observed region in the narrowest direction), making multipoles
on those scale strongly correlated \cite{tegmark1}. Thus, we binned multipoles
in bins of width $\Delta {\it l}=7 \simeq \pi/ \theta$. The signal covariance
matrix then reads:
$$
S_{ij}=\sum_b C_{{\it l}_{eff}} \sum_{{\it l} \in b}{2 {\it l} + 1 \over 4\pi}W_{\it l} P_{\it l} (\cos \theta_{ij})
$$
where we have assumed that the power spectrum is constant in each
${\it l}$--band $b$.
The sum over ${\it l}$ extends across the bin $b$, $ C_{{\it l}_{eff}}$ is
the binned power spectrum and
${\it l}_{eff}$ indicates the central value of the bands.
$P_{\it l} (\cos \theta_{ij})$ are the
Legendre polinomials and $\theta_{ij}$ is the angular separation between pixels
$i$ and $j$. $W_{\it l}$ is the beam window function.

In order to have equal area pixels in our map,
we use the icosahedron-based pixelization method proposed in \cite{tegmark2}
setting the pixel size to $\sim 4^\circ \times 4^\circ$.

Fig. \ref{spettron} shows the binned angular power spectrum $C_{{\it l}_{eff}}$ of our data
(signal plus noise) compared to the expected power spectrum (solid line)
for pure Gaussian noise $N_{\it l} = \sigma_{pix}^2 \Omega_{pix} / W_{\it l}$
with pixel variance $\sigma_{pix}^2$  and area $\Omega_{pix}$ equal to
the variance and pixel area of our map. The $1-\sigma$ error bar are derived
from the Fisher matrix. Data and theoretical noise spectra agree at
$\sim 1-\sigma$ level confirming that all the circularly polarized signals
are well buried in the noise and the validity of the Gaussian statistic up to the largest angular scales we observed.

The upper panel of Fig. \ref{spettro} displays the binned power spectrun of the
signal obtained from the maximum likelihooh method with $1-\sigma$ Fisher
matrix error bars. The values on the abscissa are
$ C_{{\it l}_{eff}} \sum_{{\it l} \in b}{2 {\it l} + 1 \over 4\pi}$
which correspond to the mean variance $\sigma_{sky}^2$ in the ${\it l}$--band
$b$. The $95 \%$ C.L. upper limits on $\sigma_{sky}^2$ are shown
in the bottom panel of Fig. \ref{spettro}
(horizontal lines donote the bin width)
while the polarization degrees in each band are summarized in Table \ref{t13}. 
Results are given for angular scale larger than the beam width
(${\it l}\lesssim 30$), higher multipoles being suppressed by the beam window
function. Further, due to the limited size of the map, lowest multipole
($\Delta {\it l} = 1-6$) can not be accurately determinated and are
disregarded.

\section{Conclusion}

We obtain $95\%$ CL upper limits to the degree of the CMB circular polarization ranging
between $5.0 \cdot 10^{-4}$ and $0.7 \cdot 10^{-4}$
at angular scales between $8^{\circ}$ and $24^{\circ}$. Results obtained with three different methods are consistent each other. 
Our observations improve the pre-existing upper
limits to the CMB circular polarization at large angular scales by an order of magnitude.
However they are still very far from
the $nK$ region where probably $V^{CMB}$ lays. Therefore they
cannot be reasonably used to set significant upper limits to primordial
magnetic field or rotation of the Universe.
\par We point out however once more that detecting CMB circular polarization
offers the possibility of detecting important features of the primeval Universe
and magnetic fields in distant clusters today studied by the
SZ effect \cite{planckSZ}. The expected signal is possibly not fainter than
the amplitude of the B-mode linear polarization that various CMB experiments
are looking for. It is therefore highly desirable that the new generations
of CMB experiments will include the
possibility of looking for circular polarization.

\par
\acknowledgments
We are grateful to the referee for helpful comments that have improved 
the manuscript.
MIPOL activity has
been supported by MIUR (Italian Ministry of University and
Research), CNR (Italian Research National Council) the
Universities of Milano and of Milano-Bicocca and the Italian
Antarctic Program (PNRA). We thanks our colleagues of Istituto of
Cosmogeofisica of CNR-Turin for hosting us and our systems at
Testa Grigia Observatory and students E. Boera, D. Colombo,
S. Cotini, L. Di Ges\'u, V.Galardo, C. Taparello, who helped us to
keep the system running at Testa Grigia in winter 2009-2010.  
L. Colombo is also thanked for useful discussions and suggestions.
\par\noindent {\it Facilities:}Testa Grigia Observatory.

\end{document}